\documentclass[twocolumn,showpacs,preprintnumbers,amsmath,amssymb,prc]{revtex4}
\usepackage{graphicx}
\usepackage{dcolumn}
\usepackage{bm}
\begin{document}
\title{First-order derivative of cluster size as a new signature of phase transition in heavy ion collisions at intermediate energies}
\author{ P. Das$^{1,2}$, S. Mallik$^{1}$, G. Chaudhuri$^{1,2}$}
\affiliation{$^1$Physics Group, Variable Energy Cyclotron Centre, 1/AF Bidhan Nagar, Kolkata 700064, India\\
$^2$Homi Bhabha National Institute, Training School Complex, Anushakti Nagar, Mumbai 400085, India}

\begin{abstract}
The phenomenon of liquid-gas phase transition occurring in heavy ion collisions at intermediate energies is a subject of contemporary interest. Phase transition is usually characterized by the specific behaviour of state variables like pressure, density, energy etc.  In heavy ion collisions there is no direct way of accessing these state variables and hence unambiguous detection of phase transition  becomes difficult. This work establishes that signatures of phase transition can be extracted from the observables which are easily accessible in experiments and these have similar behaviour as the state variables. The temperature dependence of the  first order derivative of the order parameters related to the largest and second largest cluster size (produced in heavy ion collisions) exhibit similar behavior as that of the variation of specific heat at constant volume $C_v$ which is  an established signature of first order phase transition. This motivates us to propose these derivatives as confirmatory signals of liquid-gas phase transition.  The measurement of these signals in easily feasible in  most experiments as compared to the other signatures like specific heat, caloric curve or bimodality. This temperature where the peak appears is designated to be the transition temperature and the effect of certain parameters on this has also been examined.
\end{abstract}
\maketitle
{\it{\textbf {Introduction:-}}}
The study of liquid gas phase transition in heavy ion collisions has generated a lot of interest amongst the nuclear physicists in the recent years\cite{Siemens,Gross_phase_transition,Bondorf1,Dasgupta_Phase_transition,Chomaz,Borderie2}. Different signatures of this transition have been studied extensively both theoretically \cite{Gross_phase_transition,Dasgupta_Phase_transition,Chomaz,Borderie2,Mallik10} as well as experimentally \cite{Dasgupta_Phase_transition,Chomaz,Borderie2}. First order phase transition is well characterized by some typical behaviour of different thermodynamic state variables like pressure, density, energy etc \cite{Reif,Pathria}. For example, the variation of excitation energy and specific heat with temperature are two well studied signatures theoretically in order to detect the first order phase transition\cite{Bondorf_NPA,Pochodzalla,Das2}.
The difficulty of accessing these state variables experimentally motivated us to look for more direct signatures of phase transition and in the recent papers \cite{Mallik16,Mallik20} we have established the variation of  derivative of multiplicity as a signature of liquid gas phase transition in nuclear multifragmentation. In this work we propose two new signatures of first order phase transition which can be measured more easily.  The size of the largest cluster has already been established as an order parameter for first order phase transition in heavy ion collisions. Bimodal distribution of the order parameter at a certain temperature(or excitation energy) establishes the coexistence of two phases simultaneously and well studied both theoretically and experimentally \cite{Gulminelli1,Chaudhuri_bimodality,Mallik14,Bonnet_bimodality,Mallik18}. Bimodality means two peaked distribution and the temperature where these peaks have equal height is identified as the transition temperature. There can be some ambiguity both experimentally and theoretically regarding the identification of equal heights of these peaks since the largest cluster distribution loose sharpness due to finite size of the system\cite{Mallik13}. In view of this we propose a new signature related to the largest cluster size which can be identified much easily both theoretically as well as experimentally as compared to the bimodality of the largest cluster. The temperature dependence of first-order derivative of the largest cluster display similar behaviour as that of the specific heat at constant volume. Not only that both these variables  also peak at the same temperature. We would like to emphasize that identification and determination of size of the largest cluster produced in fragmentation of hot nuclear system might be easier for the experiments as compared to the total multiplicity where is it required to detect all  the fragments produced. In this respect this proposed new signature is of much greater significance as compared to the one proposed by us recently\cite{Mallik16}. Another observable we have proposed here is related to the difference(normalized) between the sizes of the first and the second largest clusters  which also serve as an order parameter for the phase transition in nuclear multifragmentation and is well studied experimentally\cite{lefevre1,lefevre2}. The derivative of this also peaks at the same temperature as the specific heat and hence this can confirm the presence of liquid gas phase transition in nuclear multifragmentation as well. In this letter, we propose these two signatures in order to establish the existence of liquid gas phase transition in heavy ion collisions and to determine the transition temperature ar well.\\
\begin{figure}
\includegraphics[width=5.2cm,keepaspectratio=true]{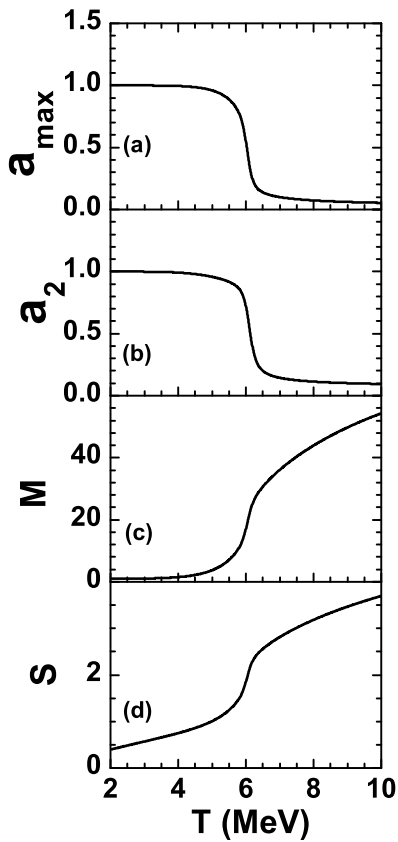}
\caption{Variation of (a) $a_{max}$, (b) $a_2$ (c) $M$ and (d) $S$ with temperature for fragmenting system of mass $A=200$.}
\end{figure}
\indent
We have used statistical models more specifically the canonical thermodynamical model(CTM)\cite{Das} in order to study the fragmentation of nuclei. In such models \cite{Das,Bondorf1,Gross1} of nuclear disassembly it is assumed that there  statistical equilibrium is attained at freeze out stage and the population of different channels of disintegration is solely decided by statistical weights in the available phase space. The calculation is done for a fixed  system size, freeze out volume and temperature.  The total multiplicity, the average size of the largest and the second largest cluster are some of the observables calculated from this model which can be measured experimentally as well. As our primary interest here is to study phase transition in nuclear system owing to the nuclear force alone,  like most theoretical models we  have considered symmetric nuclear matter where the Coulomb interaction is switched off \cite{Dasgupta,Bugaev} (the Coulomb interaction being a long range one suppresses the signatures of phase transition) and there is no distinction made between neutron and proton.\\
\indent
We give a  very brief description of the model and then present our results. Finally we will summarize and present the future outlook of this work.\\
\begin{figure}
\includegraphics[width=5.2cm,keepaspectratio=true]{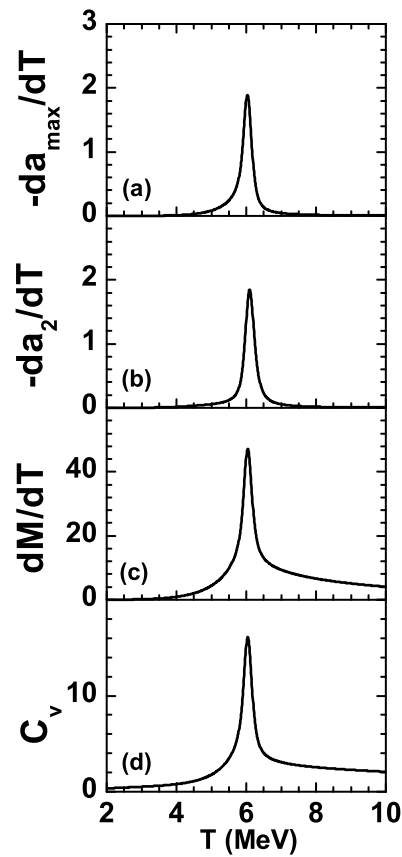}
\caption{Variation of (a) $-da_{max}/dT$, (b) $-\frac{da_2}{dT}$, (c) $\frac{dM}{dT}$ and (d) $C_v$ with temperature for fragmenting system of mass $A=200$.}
\end{figure}
\indent
{\it{\textbf {Model description:-}}}
In one component canonical model,  we consider a system of $A_0$ nucleons disintegrating at constant temperature ($T$) and freeze-out volume ($V_f$). The partitioning into different composites is done such that all partitions have the correct $A_0$. The canonical partition function is given by
\begin{equation}
Q_{A_0}=\sum\prod\frac{(\omega_{A})^{n_{A}}}{n_{A}!}
\end{equation}
\indent
Here the product is over all fragments of one break up channel and sum is over all possible channels of break-up satisfying $A_{0}=\sum A\times n_{A}$ where $n_{A}$ is the number of composites of mass number $A$ in the given channel and and $\omega_{A}$ is the partition function of the composite having $A$ nucleons. The partition function $Q_{A_{0}}$ is calculated using a recursion relation \cite{Das,Chase}

The partition function of the composite $\omega_{A}$ is a product of two parts and is given by
\begin{eqnarray}
\omega_{A}=\frac{V}{h^{3}}(2\pi mT)^{3/2}A^{3/2}\times z_{A}(int)
\end{eqnarray}
\indent
The first part is due to the translational motion and the second part $z_{A}(int)$ is the intrinsic partition function of the composite. $V$ is the volume available for translational motion. Note that $V$ will be less than $V_{f}$, the volume to which the system has expanded at break up. In general, we take $V_f$ to be equal to three to six times the normal nuclear volume.  We use $V=V_{f}-V_{0}$ , where $V_{0}$ is the normal
volume of nucleus with $A_{0}$ nucleons. The details of the model and properties of the composites used in this work are listed in details in \cite{Das}.\\
\indent
Here we introduce briefly the observables of interest in our present work, one is the average size of the largest cluster $A_{{max}}$ and other is $a_2$. Average size of the largest cluster is given as,
\begin{eqnarray}
\langle A_{max}\rangle = \sum A_{max}\,.\,Pr(A_{max})
\end{eqnarray}
where $Pr(A_{max})$ is the probability of getting a fragment of size $A_{max}$ as the largest one. This probability is given as,
\begin{eqnarray}
Pr(A_{max})=\frac{\Delta Q_{A_0}(A_{max})}{Q_{A_0} (\omega_1,\omega_2,\omega_3,...,\omega_{A_0})}
\end{eqnarray}
where,
\begin{eqnarray}
\Delta Q_{A_0}(A_{max})&=&Q_{A_0} (\omega_1,\omega_2,...,\omega_{A_{max}},0,...,0) \nonumber\\
&&-Q_{A_0} (\omega_1,\omega_2,...,\omega_{A_{max-1}},0,...,0)
\end{eqnarray}
This quantity $\Delta Q_{A_0}(A_{max})$ represents the total partition function in fragmentation of a system of size $A_0$, considering only those events where the size of the largest fragment is exactly $A_{max}$ For the suitability of this work, we will use the parameter $a_{max}= \langle A_{max} \rangle /A_0$ which is the normalized size of the largest cluster (divided by the system size).\\
\indent
The normalized variable $a_2$ is $(\langle A_{max} \rangle- \langle A_{max-1}\rangle)/(\langle A_{max} \rangle+ \langle A_{max-1} \rangle)$ \cite{lefevre1,lefevre2},
where $\langle A_{max-1} \rangle$ is the average size of the second largest fragment. One can calculate it, by proceeding in a similar way \cite{Chaudhuri_bimodality} of ${\langle A_{max} \rangle}$. Thus if $Pr_2(A_{max-1})$ is the probability for $A_{max-1}$ to be the second largest fragment size, then
\begin{eqnarray}
\langle A_{max-1} \rangle = \sum A_{max-1}\,.\,Pr_2(A_{max-1})
\end{eqnarray}
\indent
Now, to get this probability, we see that $A_{max-1}$ can be the second largest if (a) there is at least one fragment of size $A_{max-1}$ and just one fragment of size $A_{max} > A_{max-1}$ or if (b) there are more than one fragment of size $A_{max-1}$ but no fragment larger than it; $A_{max}=A_{max-1}$
The partition function for the case (a) is
\begin{eqnarray}
Q_a=\sum \omega_{A_{max}}\, . \, \Delta Q_{{A_0}-{A_{max}}}(A_{max-1})
\end{eqnarray}
where the sum goes from $(A_{max-1}+1)$ to its maximum possible value and for the case (b) is
\begin{eqnarray}
Q_b&=&\Delta Q_{A_0}(A_{max-1})-\omega_{A_{max-1}} \nonumber\\
&&\,.\, Q_{A_0-A_{max-1}}(\omega_1,\omega_2,...,\omega_{{A_{max-1}}-1},0,....)
\end{eqnarray}
The first term is the total partition function for the channels where the largest cluster size is $A_{max-1}$ but the number of such clusters can be one or more. The second term gives the total partition function for the channels where the number of fragments of size $A_{max-1}$ (i.e., largest cluster) is just one. So the difference is the the partition function for case (b).
Therefore, the second largest cluster probability will be,
\begin{eqnarray}
Pr_2(A_{max-1})=[Q_a + Q_b]/Q_{A_0}
\end{eqnarray}
Once we get the probability, using Eqn.(6) ${\langle A_{max-1} \rangle}$ can be calculated.\\
\begin{figure}
\includegraphics[width=5.4cm,keepaspectratio=true]{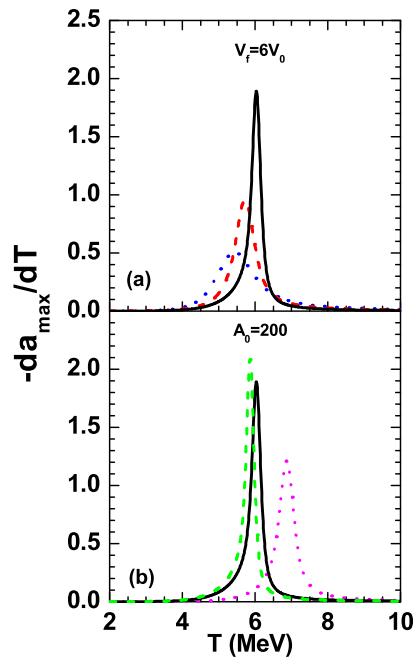}
\caption{(Color Online) Variation of $-da_{max}/dT$ with temperature (a) at constant freeze-out volume $V_f=6V_0$ but for there fragmenting systems of mass 50 (blue dotted line), 100 (red dashed line) and 200 (black solid line) and (b) for same fragmenting system of mass 200 but at three constant freeze-out volumes $V_f=2V_0$ (magenta dotted line), $V_f=6V_0$ (black solid line)and $V_f=8V_0$ (green dashed line).}
\end{figure}
\indent
{\it{\textbf {Results:-}}}
The size of the largest cluster formed in the fragmentation of an excited nuclei behaves as an order parameter for first order phase transition\cite{Gulminelli1,Krishnamachari,Pleimling,Chaudhuri_largest_cluster}. The  largest cluster size varies (decreases) very slowly as the temperature rises and then suddenly as the liquid-gas transition temperature is reached, there is a sudden fall in the largest cluster size after which it again decreases very slowly. This behaviour is depicted in Fig. 1(a). $a_2$ which represents the normalized size difference of the first and the second largest cluster also displays similar behaviour as $a_{max}$ and that is shown in Fig. 1(b). This parameter is also markedly different in the liquid and in the  gas phases and can be considered to be an order parameter of the transition. Fig1(c) and (d) depicts the change in total multiplicity and entropy per nucleon ($S$) as temperature is increased and similar behaviour is noticed. The sudden decrease(or increase) of all these four quantities occur at almost the same temperature which is the transition temperature and in this case it is about 6 MeV. In our last work\cite{Mallik16} in this subject, we have established the multiplicity derivative as a signature for first order phase transition. Similar (sudden rise or fall) behaviour of the size of the largest cluster and also that of $a_2$ motivated us to investigate the behaviour of the derivative of the same. This is plotted in Fig 2(a) and (b).  It shows that the magnitude of the derivative of $a_{max}$ and $a_2$ displays a pattern quite similar to that of specific heat per nucleon, $C_v$ (Fig. 2(d)) and that of $dM/dT$ (Fig. 2(c)). The derivative shows a peak at the same transition temperature which is indicative of liquid gas phase transition. This signature is much easier to access both theoretically and experimentally as compared to the bimodality in the probability distribution of the largest cluster. The later has been used  so far in order to detect the existence of phase transition in nuclear multifragmentation but to detect two peaks(bimodality) of equal height in a distribution at a particular temperature (or excitation energy) is far more a difficult job than to simply calculate the derivative in its size with temperature or excitation energy. We strongly believe that this new proposed signature related to the largest cluster size will definitely provide a great impetus to the study of liquid gas phase transition in heavy ion collisions. It can be accessed easily in most experiments as compared to all other standard phase transition signals. In this context we would like to point out that the distributions of $\frac{da_{max}}{dT}$ and $\frac{da_2}{dT}$ might lose their sharpness to some extent after the evaporation from the hot fragments. Also the fission process which might be dominant at lower temperature in case of heavy nuclei can distort the signature to some extent if Coulomb interaction is included.\\
\begin{figure}[t]
\includegraphics[width=5.4cm,keepaspectratio=true]{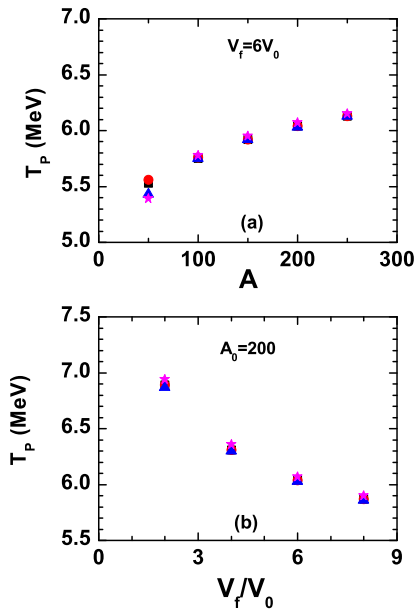}
\caption{(Color Online) Dependence of the peak position of $-da_{max}/dT$, $-\frac{da_2}{dT}$,  $\frac{dM}{dT}$ and $C_v$ on fragmenting system size (upper panel) and freeze-out volume (lower panel).}
\end{figure}
\indent
Next we show the dependence of this transition temperature on the size of the fragmenting source. Fig 3.(a) depicts this change for $a_{max}$. The calculations are done for three fragmenting source size of 50, 100 and 200 and it is observed that the transition temperature decreases as the source size is decreased. This implies that smaller system fragments more easily and at a lower temperature as compared to its bigger counterparts. Also the peak becomes more sharper that is height increases as the system size increases which establishes that phase transition signatures are better visible in larger systems.  We also examine the effect of freeze-out volume on the transition temperature and this is displayed in Fig. 3.(b). The calculations are done for three freeze out volumes and it is seen that more is the freeze out volume, less is the transition temperature. This is what we expect because higher volume(lower density) will favour the disintegration of the nucleus resulting in lower transition temperature. The effect of freeze-out volume and source size has also been examined for the parameter $a_2$ and exactly similar behavior has been noticed and hence we have not presented those results here for brevity.\\
\indent
In the last figure(Fig. 4) we have plotted the transition temperature as function of source size (upper panel) for a fixed freeze-out volume and the freeze-out volume (lower panel) for a fixed source. Here we have shown the variation of peak position for $da_{max}/dT$, $da_2/dT$, $dM/dT$ and last but not the least $C_V$ in both the figures. The results for all the variables coincided with each other and this further establishes our claim for $da_{max}/dT$, $da_2/dT$ to be the signatures of the phase transition. The small difference between the four signatures is attributed to the finite size of the system.\\
\indent
{\it{\textbf {Summary and future outlook:-}}}
The variation of the derivative of the largest cluster size as well as that of $a_2$ (normalized size difference between largest and the second largest cluster) with temperature has been proposed as the signals for detection of  phase transition in heavy ion collisions.
These new signatures  will surely provide some definite answer to the long standing problem of extraction of experimental signals for phase transition in heavy ion collisions at intermediate energies.  They display exactly similar behaviour as specific heat and hence can be claimed to be the confirmatory tests for occurrence of nuclear liquid gas phase transition. The size of the largest and the second largest cluster can be easily measured in most of the laboratories at different energies and this will suffice in analysing the phase transition unambiguously in contrary to most of the signals that are in use so far.  The transition temperature can be located with much more precision as compared to other signals  from the peak position of the derivatives. The extension of this work to nuclei with two kinds of particles is in progress.\\

\end{document}